\documentstyle[12pt]{article}

\advance \textheight by \topskip
\def\A#1{\raisebox{.2ex}{$\displaystyle
    \mathop{A}^{\scriptscriptstyle [#1]}$}{}}
\def\e#1{\raisebox{.2ex}{$\displaystyle
    \mathop{e}^{\scriptscriptstyle [#1]}$}{}}

\makeatletter
\@addtoreset{equation}{section}

\makeatother

\begin{document}

\begin{titlepage}
\hbox to \hsize{\hfil TUM--HEP--277/97}
\hbox to \hsize{\hfil NF/DF--06/97}
\hbox to \hsize{\hfil May 1997}
\vfill
\large \bf
\begin{center}
P,T-invariant system of Chern-Simons fields:\\
Pseudoclassical model and hidden symmetries
\end{center}
\vskip1cm
\normalsize
\begin{center}
Khazret S. Nirov${}^{a,}${}{\footnote{Alexander von Humboldt fellow;
on leave from the {\it Institute for Nuclear Research, Moscow,
Russia}}${}^,$\footnote{E--mail: nirov@dirac.physik.uni-bonn.de}} $\,$ 
~and~ Mikhail S. Plyushchay${}^{b,c,}${}\footnote{E--mail: 
plyushchay@mx.ihep.su}\\
{\small \it ${}^a{}$Institut f\"ur Theoretische Physik T30,
 Physik Department,}\\
{\small \it Technische Universit\"at M\"unchen,
 D-85747 Garching, Germany}\\
{\small \it ${}^b{}$Departamento de Fisica --- ICE, 
 Universidade Federal de Juiz de Fora}\\
{\small \it 36036-330 Juiz de Fora, MG Brazil}\\
{\small \it ${}^{c}{}$Institute for High Energy Physics, 
Protvino, Moscow region, 142284 Russia}
\end{center}
\vskip2cm
\begin{abstract}
\noindent 
We investigate hidden symmetries of $P$,$T$-invariant system
of topologically massive U(1) gauge fields. For this purpose,
we propose a pseudoclassical model giving rise to this field 
system at the quantum level. The model contains a parameter, 
which displays a quantization property at the classical and 
the quantum levels and demonstrates a nontrivial relationship 
between continuous and discrete symmetries. Analyzing the
integrals of motion of the pseudoclassical model, we identify 
U(1,1) symmetry and S(2,1) supersymmetry as hidden symmetries 
of the corresponding quantum system. Representing the hidden 
symmetries in a covariant form, we show that one-particle 
states realize an irreducible representation of a 
non-standard super-extension of the $(2+1)$-dimensional 
Poincar\'e group labelled by the zero eigenvalue of the 
superspin.


\end{abstract}
\vfill
\end{titlepage}

\section{Introduction}

Topologically massive gauge fields \cite{top}, originated from the
$\theta$-vacuum of four-dimensional theories \cite{theta}, turned 
out to be the basic tool for constructing models which possess quantum 
Hall effect \cite{hall} and high-temperature superconductivity 
\cite{T-sc}. Actually, considerable interest in 3d field theories
is highly motivated by perspectives they give us for better
understanding critical phenomena generic to 4d physics \cite{3D4D}. 
In this, a particular position of the high-temperature superconductivity 
is cogent: high-$T_c$ superconducting materials have quasi-planar 
structures \cite{quasi}, so that they can effectively be described by 
three-dimensional models.

We deal here with the simplest $P$,$T$-invariant system 
of Chern-Simons vector U(1) gauge fields, given in terms
of a self-dual free massive field theory \cite{TPN}. 
The corresponding source-free equations are first order 
differential equations 
${\cal L}^\epsilon_{\mu\nu} {\cal F}^\nu_\epsilon = 0$,
where 
${\cal L}^\epsilon_{\mu\nu} \equiv 
(i \varepsilon_{\mu\nu\lambda} P^\lambda + \epsilon m \eta_{\mu\nu})$,
$P_\mu = -i \partial_\mu$,  $\eta_{\mu\nu} = diag(-1,+1,+1)$, 
$\epsilon = +$ or $-$, 
and the totally antisymmetric tensor 
$\varepsilon^{\mu\nu\lambda}$ 
is normalized by 
$\varepsilon^{012} = 1$. 
Due to the basic equations, the field 
${\cal F}^\mu_\epsilon$ 
satisfies also Klein-Gordon equation
$(P^2 + m^2) {\cal F}^\mu_\epsilon = 0$
and the transversality condition
$P_\mu {\cal F}^\mu_\epsilon = 0$. 
It is clear from the definition above that ${\cal F}^\mu_\epsilon$ 
carries massive irreducible representation of spin $s = -\epsilon 1$ 
of the 3d Poincar\'e group. As was shown in Ref. \cite{DJ}, this
formulation of the theory is essentially equivalent to the original 
one \cite{top}. 

Already in pioneering works \cite{top} it was noted that topological 
mass terms are odd under the parity and time-reversal transformations, 
and the full set of discrete $C$, $P$ and $T$ symmetries may be restored 
if one doubles the number of fields and introduces opposite sign mass 
terms. In the case under consideration, when taking the action 
\begin{equation}
{\cal A} = \int d^3x \left( 
{\cal F}^\mu_+ {\cal L}^+_{\mu\nu} {\cal F}^\nu_+ 
+ {\cal F}^\mu_- {\cal L}^-_{\mu\nu} {\cal F}^\nu_- \right),
\label{calA}
\end{equation} 
we get $P$,$T$-invariant system of topologically massive vector 
U(1) gauge fields \cite{top}. This observation plays an essential
role in constructing models of high-temperature superconductors.
Actually, single spin state models predict observable parity and
time-reversal violation in corresponding superconductors \cite{T-inv},
for which experiments still give no evidence \cite{exp}. Besides, 
the problem of cancellation between single bare and radiatively 
generated Chern-Simons terms \cite{Red} arises in the conventional 
models \cite{cancel}. For these fundamental reasons, it is desirable 
to have parity and time-reversal conserving system modelling 
high-$T_c$ superconductors without, at least, these serious 
obstructions \cite{T-inv-mod}. For the same reasons, we shall
pay particular attention to the requirement of these discrete 
symmetries.

The relevance of 3d field theories to critical phenomena might 
be explained by some hidden symmetries. From this point of view,
it is of interest to investigate properties of parity and
time-reversal conserving systems which are considered to be 
relevant to high-temperature superconductivity. Such a program
has been realized for planar fermions \cite{GPS}, where the
authors elucidated a rich set of hidden symmetries. The results
of Ref. \cite{GPS} were obtained from analysis of the pseudoclassical
model of a relativistic spinning particle proposed in Ref. \cite{CPV}. 
Classical particle models are indeed useful for clarifying problems 
of more complicated quantum mechanical and field systems and for 
revealing hidden properties of corresponding quantum systems and 
understanding their nature. 

In this paper we propose a pseudoclassical model by means of 
which we analyze $P$,$T$-invariant system of Chern-Simons fields.
The model contains a $c$-number parameter at a mass term for spin 
variables. This one, which we call the model parameter, displays 
a quantization property both at the classical and quantum levels. 
Although the variation of the parameter does not affect the 
discrete space-time symmetries of the pseudoclassical model, 
its values are crucial for continuous global symmetries. 
Actually, there are special discrete values of the model 
parameter at which the system has a maximal number of 
integrals of motion. The same values of the parameter 
turn out to be special quantum mechanically: they are 
separated by the requirement of maximality of global
symmetry of the physical state space at the quantum level.
Moreover, we shall see that only at these special values
discrete parity and time-reversal symmetries are conserved
in the corresponding quantum theory. This result does actually 
indicate a profound relationship between discrete and continuous
global symmetries. When considering algebras of the integrals
of motion, we shall elucidate hidden U(1,1) symmetry and S(2,1)
supersymmetry of the $P$,$T$-invariant system of Chern-Simons
U(1) gauge fields. We shall also demonstrate that this system 
realizes an irreducible representation of a non-standard
super-extension of the $(2+1)$-dimensional Poincar\'e group,
namely $\rm ISO(2,1|2,1)$. The non-standard character of this
supergroup means, in particular, that, unlike the usual 
supersymmetries, the anticommutator of corresponding 
supercharges results in an operator different from the 
Hamiltonian of the system. 

It is interesting to observe that non-standard supersymmetries 
and quantization of parameters turn out to be generic to systems 
with nontrivial topology of the corresponding configuration or 
phase spaces. For instance, when investigating space-time symmetries 
in terms of motion of pseudoclassical spinning point particles 
\cite{spin1}--\cite{spin4}, Gibbons, Rietdijk and van Holten 
\cite{non-st1} elucidated the existence of a non-standard 
supersymmetry (see also Ref. \cite{non-st2}). 
In this, the Poisson brackets of the odd Grassmann generators 
give rise to an even integral of motion different from the 
Hamiltonian of the system. A non-standard supersymmetry with 
the same feature appeared in studying hidden symmetries 
\cite{hid-mono} of a 3d monopole \cite{mono}. And the 
quantization of the dimensionless mass-coupling-constant ratio 
of the non-Abelian 3d vector fields \cite{top,mono}, discovered 
by Deser, Jackiw and Templeton, is caused by nontrivial homotopy 
properties of these fields. 

The paper is organized as follows. In Section 2 the pseudoclassical
model and its Lagrangian symmetries are described. Section 3 is
devoted to Hamiltonian description of the model. In this Section
the solutions to the equations of motion and the set of the integrals
of motion are presented. In Section 4 the corresponding quantum
theory is constructed. Provided that, the hidden (super)symmetries
of the system under consideration are revealed. Covariantization of 
the symmetry relations is performed in Section 5. After Concluding 
remarks, some useful from a technical point of view formulas are 
gathered in the Appendix.

Everywhere in the text repeated indices imply the corresponding
summation.

\section{The pseudoclassical model and its symmetries}

\subsection{The action}

The pseudoclassical model we are going to analyze here 
is given by the action
\begin{equation}
A_q = \int^{\tau_f}_{\tau_i} L_q d\tau + \Gamma_\xi, 
\label{Action}
\end{equation}
where $L_q$ is the Lagrangian
\begin{equation} 
L_q = \frac{1}{2e}\left(\dot{x}_\mu 
- \frac{i}{2} v \varepsilon_{\mu\nu\lambda}
\xi^\nu_a \xi^\lambda_a \right)^2 - \frac{1}{2}e m^2 
- i q m v \xi_1^\mu\xi_{2\mu}
+ \frac{i}{2} \xi_a^\mu \dot{\xi}_{a\mu}, 
\label{L}
\end{equation}
and $\Gamma_\xi$ means a boundary term, 
$\Gamma_\xi = \frac{i}{2}\xi^\mu_a(\tau_f) \xi_{a\mu}(\tau_i)$. 
The configuration space of the system is described by the set 
of variables $x_\mu$, $\xi_a^\mu$, $a=1,2$, $e$ and $v$.
In this, $x_\mu$, $\mu = 0,1,2$, denote space-time coordinates 
of the particle, $\xi^\mu_a$ are real Grassmann odd variables 
forming two Lorentz vectors, $e$ and $v$ are even Lagrange 
multipliers, and $q$ is a real $c$-number parameter. 

The presence of the boundary term is caused by the form of the equations
of motion for the Grassmann variables which are differential equations
of the first order \cite{GSHT}. The action $A_q$ is extremal on the
trajectories satisfying the boundary conditions
$\delta \xi_a^\mu(\tau_i) + \delta \xi_a^\mu(\tau_f) = 0$.

\subsection{Discrete symmetries}

The system given by Eqs. (\ref{Action})-(\ref{L}) is invariant 
under the discrete parity and time-reversal transformations
\begin{equation}
P : X^\mu \rightarrow \tilde\varepsilon(X^0, -X^1, X^2),
\qquad
T : X^\mu \rightarrow \tilde\varepsilon(-X^0, X^1, X^2),
\label{PT1}
\end{equation}
where $X^\mu = x^\mu, \xi_1^\mu, \xi_2^\mu$, and
\begin{equation}
P :, T : E \rightarrow \tilde\varepsilon E,
\end{equation}
where $E = e, v$. In these discrete symmetry transformation
laws $\tilde\varepsilon = +$ for the vector $x_\mu, \xi_1^\mu$
and scalar $e$ variables, and $\tilde\varepsilon = -$ for 
$\xi_2^\mu$ and $v$ implying that $\xi_2^\mu$ is a pseudovector
and $v$ is a pseudoscalar.

One of the most important features of our pseudoclassical model is 
that in the classical theory parity and time-reversal invariance 
take place for any value of the parameter $q$. Nevertheless, we 
shall see that the case of $|q| = 2$ is particular both at the 
classical and the quantum levels of the theory, and that the 
quantization of the model (\ref{Action})-(\ref{L}) results 
in the $P$,$T$-invariant system of topologically massive 
vector U(1) gauge fields. 

\subsection{Global symmetries}

In addition to the Poincar\'e invariance, 
the action (\ref{Action}) is invariant against 
the following set of global transformations:
\begin{equation}
\delta_{\lambda} \xi_{a\mu} = \lambda \epsilon_{ab} \xi_{b\mu},
\label{lambda}
\end{equation}
where $\epsilon_{ab} = -\epsilon_{ba}$, $a,b = 1,2$, $\epsilon_{12} = 1$,
\begin{eqnarray}
\delta_{\nu} x_\mu &=& \nu \epsilon_{ab} 
\xi_{a\mu} \xi_{b\lambda} p^\lambda, \label{n1}\\
\delta_{\nu} \xi_{a\mu} &=& -i \nu \epsilon_{ab} 
p_\mu p_\lambda \xi_b^\lambda,
\label{n2}
\end{eqnarray}
and
\begin{eqnarray}
\delta_\theta x_\mu &=& i\theta \varepsilon_{\mu\nu\lambda} 
\xi_a^\nu \xi_a^\lambda, \label{t1}\\
\delta_{\theta} \xi_{a\mu} &=& -2\theta \varepsilon_{\mu\nu\lambda} 
p^\nu \xi_a^\lambda, \label{t2}
\end{eqnarray}
where we have introduced the notation
$p_\mu = e^{-1} \left( \dot{x}_\mu - 
\frac{i}{2} v \varepsilon_{\mu\nu\lambda}
\xi_a^\nu \xi_a^\lambda \right)$.
In these transformations $\lambda$, $\nu$ and $\theta$ are 
even constant infinitesimal parameters. Further we shall 
find the corresponding generators of these global symmetries.

\subsection{Local symmetries}

The system has two local symmetries. One of them is a 
reparametrization invariance defined with respect to
the transformations 
\begin{equation}
\delta_\alpha E = \frac{d}{d\tau} (\alpha E), \qquad
\delta_\alpha X = \alpha \dot X, 
\end{equation}
where $E = e, v$ and $X = x_\mu, \xi_{a\mu}$, 
changing the Lagrangian $L_q$ by
$\delta_\alpha L_q = \frac{d}{d\tau} (\alpha L_q)$.
As a consequence, the corresponding action 
is extremal if the boundary conditions
$\alpha(\tau_i) = \alpha(\tau_f) = 0$
for the infinitesimal gauge parameter $\alpha$ are fulfilled.

Another local symmetry transformation is of the form
\begin{eqnarray}
\delta_\beta x_\mu &=& \frac{i}{2} \beta \varepsilon_{\mu\nu\lambda}
\xi^\nu_a \xi^\lambda_a, \\
\delta_\beta \xi_{a\mu} &=& -\beta \left( \varepsilon_{\mu\nu\lambda}
p^\nu \xi_a^\lambda - q m \epsilon_{ab} \xi_{b\mu} \right), 
\label{2.12} \\
\delta_\beta v &=& \dot \beta.
\end{eqnarray}
For the Lagrangian $L_q$ we obtain 
$\delta_\beta L_q = \frac{d}{d\tau} \left( \frac{i}{2} 
\beta \varepsilon_{\mu\nu\lambda} p^\mu
\xi_a^\nu \xi_a^\lambda \right)$, 
so that the action is invariant provided that 
the boundary conditions
$\beta(\tau_i) = \beta(\tau_f) = 0$
on the gauge parameter $\beta$ are imposed.

It is interesting to note here that if we formally relate 
global and local symmetry parameters as
$\lambda = 2 q m \theta = q m \beta$,
we get the local $\beta$-symmetry to be a superposition of $\lambda$ 
and $\theta$ global symmetries,
\begin{equation}
\delta_\beta = \delta_\lambda + \delta_\theta. 
\label{gl-loc}
\end{equation}
Having a Hamiltonian description of the system we shall
obtain the generators of the local symmetry transformations 
and shall reveal the origin of the property (\ref{gl-loc}).

\section{Hamiltonian description of the model}

\subsection{Canonical structure and constraints}

Let us construct the Hamiltonian description of the
model. The nontrivial Poisson-Dirac brackets following 
from the Lagrangian (\ref{L}) are
\begin{equation}
\{x_\mu,p_\nu\} = \eta_{\mu\nu}, \qquad 
\{\xi_a^\mu,\xi_b^\nu\} = - i \delta_{ab} \eta^{\mu\nu}, 
\end{equation}
\begin{equation}
\{e,p_e\} = 1, \qquad \{v,p_v\} = 1.
\end{equation}
The model possesses two sets of primary, 
$p_e \approx 0, p_v \approx 0$, 
and secondary, 
\begin{equation}
\phi = \frac{1}{2} (p^2 + m^2) \approx 0, \quad
\chi = \frac{i}{2} \left(\varepsilon_{\mu\nu\lambda} p^\mu
\xi^\nu_a \xi^\lambda_a + q m \epsilon_{ab} \xi_a \xi_b\right) 
\approx 0,
\label{sec}
\end{equation}
constraints forming the trivial algebra of the first class 
with respect to the above brackets. As a consequence of 
the reparametrization invariance, the Hamiltonian of 
our model is a linear combination of the constraints:
\begin{equation}
H = e\phi + v\chi + u_1 p_e + u_2 p_v, 
\end{equation}
with the coefficients at the primary constraints being 
arbitrary functions of the evolution parameter $\tau$. 
It is easy to see that the reparametrization invariance 
is generated by the set of the constraints 
$p_e \approx 0$ and $\phi \approx 0$,
while the generators of the local $\beta$-symmetry are
the constraints $p_v \approx 0$ and $\chi \approx 0$
\cite{GSHT}.

\subsection{Equations of motion}

Essential equations of motion of the system are
\begin{eqnarray}
\dot{p}_\mu &=& 0, \\
\dot{x}_\mu &=& e p_\mu + \frac{i}{2}
v \varepsilon_{\mu\nu\lambda} \xi^\nu_a \xi^\lambda_a, \\
\dot{\xi}_{a\mu}
&=& - v (\varepsilon_{\mu\nu\lambda} p^\nu \xi^\lambda_a 
- q m \epsilon_{ab} \xi_{b\mu}). \label{3.7}
\end{eqnarray}
{}From these equations we immediately find that the 
energy-momentum vector 
$p_\mu$
and the total angular momentum vector 
${\cal J}_\mu = -\varepsilon_{\mu\nu\lambda} x^\nu p^\lambda 
+ \frac{i}{2} \varepsilon_{\mu\nu\lambda} \xi^\nu_a \xi^\lambda_a$ 
are integrals of motion. 
Since Eqs. (\ref{3.7}) are generated by the nilpotent
constraint only, $\dot{\xi}_{a\mu}=v\{\xi_{a\mu},\chi\}$,
it can be considered as a Hamiltonian of the spin variables.

The equations of motion for the Lagrange multipliers 
are not important for our analysis, therefore we shall
not write them down.

To solve the equations for the spin variables $\xi^\mu_a$, 
it is convenient to use complex mutually conjugate odd variables
$b^{\pm}_\mu = \frac{1}{\sqrt{2}} (\xi_{1\mu} \pm i \xi_{2\mu})$
with nontrivial brackets
$\{b^+_\mu,b^-_\nu\} = - i\eta_{\mu\nu}$.
The new odd variables satisfy the equations
\begin{equation}
\dot{b}^\pm_\mu = 
- v \left( \varepsilon_{\mu\nu\lambda}p^\nu 
\pm i q m \eta_{\mu\lambda} \right) b^{\pm \lambda}.
\end{equation}

Taking into account the mass-shell constraint, 
we introduce the general notation 
$f^{(\alpha)} \equiv f^\mu e^{(\alpha)}_\mu$ 
for the projection of any Lorentz vector $f^\mu$ 
onto the complete oriented triad
$e^{(\alpha)}_\mu(p)$, $\alpha = 0,1,2,$ 
defined by the relations
\begin{equation}
e^{(0)}_\mu = \frac{p_\mu}{\sqrt{-p^2}}, \quad 
e^{(\alpha)}_\mu \eta_{\alpha\beta} e^{(\beta)}_\nu = \eta_{\mu\nu},
\quad
\varepsilon_{\mu\nu\lambda} e^{(0)\mu} e^{(i)\nu} e^{(j)\lambda} =
\varepsilon^{0ij}.
\end{equation}
It is important to note here that $e^{(i)}_\mu(p)$ are not Lorentz 
vectors, and so, the projections $f^{(i)}$ of a Lorentz vector 
$f_\mu$ onto these triad components are not covariant quantities, 
while $f^{(0)}$ is a Lorentz scalar \cite{MP}.

In terms of these, we find that the odd spin variables have 
the following evolution law:
\begin{equation}
b^{(0)\pm}(\tau) = e^{\mp i q \omega(\tau)} b^{(0)\pm}(\tau_i), 
\end{equation}
\begin{equation}
b^{(i)\pm}(\tau) = e^{\mp i q \omega(\tau)} 
\left[ \cos\omega(\tau) b^{(i)\pm}(\tau_i) 
+ \varepsilon^{0ij} \sin\omega(\tau) b^{(j)\pm}(\tau_i) \right], 
\end{equation}
with 
$\omega(\tau) \equiv \omega(\tau;\tau_i) 
= m \int_{\tau_i}^\tau v(\tau^\prime) d\tau^\prime$,
so that $b^{(0)\pm}$ are harmonic-like variables, while the solution 
for $b^{(i)\pm}$ includes an additional SO(2) rotation.

In terms of the initial odd variables $\xi^\mu_a$ the solutions
to the equations of motion can be written as follows:
\begin{eqnarray}
\xi_{a\mu}(\tau) &=& g_{\mu\nu}(\tau)\biggl(
\xi_a^\nu(\tau_i)\cos q\omega(\tau) + \epsilon_{ab} 
\xi_b^\nu(\tau_i)\sin q\omega(\tau) \biggr), \quad a = 1,2, \\
x_\mu(\tau) &=& p_\mu \int^\tau_{\tau_i} e(\tau^\prime) d\tau^\prime
- \frac{1}{2m} e^{(0)}_\mu \pi_{\nu\lambda}
\xi_1^\nu(\tau) \xi_2^\lambda(\tau)
+ \frac{i}{m}\xi^{(0)}_a(\tau)\xi_{a\mu}(\tau) + x_\mu(\tau_i),
\label{x}
\end{eqnarray} 
where we have introduced the notations
\begin{equation}
g_{\mu\nu}(\tau) = -e^{(0)}_\mu e^{(0)}_\nu 
+ \pi_{\mu\nu} \cos\omega(\tau) 
+ \varepsilon_{\mu\nu\lambda} e^{(0)\lambda}\sin\omega(\tau), 
\qquad
\pi_{\mu\nu} = \eta_{\mu\nu} + e^{(0)}_\mu e^{(0)}_\nu.
\label{pi}
\end{equation}
We see that for $q \neq 0$ the evolution mixes the initial data 
of the spin variables. The terms with the odd variables in the 
solution for the coordinates of the particle (\ref{x}) describe 
the pseudoclassical analog of the quantum Zitterbewegung 
\cite{Dirac,JM,spin2,spin3,MP}.

We have seen also that it is quite natural to use complex variables
$b^\pm_\mu$ instead of their real and imaginary parts $\xi_{a\mu}$,
and so, in what follows, we will do work in terms of these complex
spin variables. 

\subsection{Integrals of motion}

{}From the solutions to the equations of motion we obtain quadratic 
nilpotent integrals of motion 
\begin{equation}
{\cal N}_0 = b^{(0)+} b^{(0)-}, \qquad 
{\cal N}_\perp = b^{(i)+} b^{(i)-}, \qquad
{\cal S} = i\varepsilon^{0ij} b^{(i)+} b^{(j)-} 
\equiv {\cal J}^{(0)}.
\end{equation}
The integral of motion ${\cal N}_0$ is the generator of
the global $\nu$-symmetry transformation (\ref{n1}),(\ref{n2}).
The global $\lambda$-symmetry transformation (\ref{lambda}) is 
generated by the combination 
${\cal N} = - {\cal N}_0 + {\cal N}_\perp$.
The global SO(2) rotations with the parameter $\theta$ 
(\ref{t1}),(\ref{t2}) are generated by the integral of motion
${\cal S}\sqrt{-p^2}$. 
We see that on the mass shell the constraint function 
$\chi$ can be represented as a linear combination 
of the quadratic integrals of motion,
$\chi = m\left( {\cal S} - q {\cal N} \right)$,
and so, regarding this nilpotent constraint as the generator of
the gauge $\beta$-transformation and the integrals of motion as
the generators of global symmetry transformations, we see the  
reason of the above mentioned relation (\ref{gl-loc}) of global 
and local symmetries of the model.

The case of $q=0$ is dynamically degenerated with the
variables $b^{(0)\pm}$ being trivial integrals of motion,
$b^{(0)\pm}(\tau) = b^{(0)\pm}(\tau_i)$. In this case the 
constraint $\chi$ generates only SO(2) rotations, which do
not transform variables $b^{(0)\pm}$. As we shall see, this 
special case is completely excluded on the quantum level.

Further, we have the nilpotent second order quantities
\begin{equation}
B^\pm = 
\left( b^{(2)+} b^{(2)-} - b^{(1)+} b^{(1)-} \right) \pm
i \left( b^{(2)+} b^{(1)-} + b^{(1)+} b^{(2)-} \right)
\end{equation}
satisfying a simple evolution law 
$\dot{B}^\pm = \pm 2imv B^\pm$
with an obvious harmonic-like solution
$B^\pm(\tau) = e^{\pm 2i\omega(\tau)} B^\pm(\tau_i)$.
Recalling the evolution law for the odd variables $b^{(0)\pm}$
we obtain that if and only if $|q| = 2$, there are two 
additional third order nilpotent integrals of motion 
in the model, namely
\begin{equation}
{\cal B}^\pm_+ = B^\pm b^{(0)\pm}, \qquad 
{\cal B}^+_+ = ({\cal B}^-_+)^* \qquad
{\rm for} ~~q = 2, 
\end{equation}
or
\begin{equation}
{\cal B}^\pm_- = B^\pm b^{(0)\mp}, \qquad 
{\cal B}^+_- = ({\cal B}^-_-)^* \qquad
{\rm for} ~~q = - 2,
\end{equation} 
which are {\it local} in the evolution parameter $\tau$ quantities.

The quadratic integrals of motion ${\cal N}_0$, ${\cal N}_\perp$ 
and ${\cal S}$ form trivial algebra with respect to the  
canonical structure
$\{b^{(\alpha)+},b^{(\beta)-}\} = -i\eta^{\alpha\beta}$. 
Besides, we find that
\begin{equation}
{\cal N}^2_0 = 0, \qquad {\cal N}_\perp {\cal S} = 0, \qquad
{\cal N}^2_\perp = - {\cal S}^2.
\end{equation}
We have also for $q=2$ (the case of $q=-2$ can easily be 
reproduced by the change $\xi_1^\mu \leftrightarrow \xi_2^\mu$):
\begin{eqnarray}
&{\cal B}^+_+ {\cal B^-_+} = 2 {\cal N}_0 {\cal N}^2_\perp,&
\label{b1}\\
&\{{\cal B}^+_+,{\cal B}^-_+\} = -2i {\cal H}_+, \qquad
\{{\cal B}^\pm_+,{\cal H}_+\} = 0,& 
\label{b2}\\
&\{{\cal B}^\pm_+,{\cal N}_0\} = \mp i {\cal B}^\pm_+, \qquad
\{{\cal B}^\pm_+,{\cal N}_\perp\} = 0, \qquad
\{{\cal B}^\pm_+,{\cal S}\} = \pm 2i {\cal B}^\pm_+,&
\label{b3}
\end{eqnarray}
where 
${\cal H}_+ = {\cal N}^2_\perp + 2{\cal N}_0 {\cal S}$. 

For arbitrary $q > 0$ or $q < 0$ one can construct 
{\it nonlocal} integrals of motion
${\cal B}^\pm_{+q} = B^\pm b^{(0)\pm} e^{\pm i(q-2)\omega(\tau)}$
and
${\cal B}^\pm_{-q} = B^\pm b^{(0)\mp} e^{\mp i(q+2)\omega(\tau)}$,
respectively.
For the particular values of the model parameter, $q = \pm 2$,
these quantities become local in $\tau$, coinciding with the
integrals of motion ${\cal B}^\pm_\pm$. These integrals generate
global symmetry transformations, acting on the canonical variables 
$X = x_\mu, b^\pm_\mu$ as 
$\delta_\pm X = \gamma_\pm \{X,{\cal B}^\pm_\pm\}$,
where $\gamma_\pm$ are corresponding odd constant
infinitesimal transformation parameters.

Thus, here we have observed some phenomenon of 
{\it classical quantization}: there are two special 
values of the parameter $q$, $q = \pm 2$, when, and
only when, the system has additional (local in $\tau$) 
nontrivial integrals of motion. These integrals are the 
generators of corresponding global symmetry transformations, 
and so, the system has maximal global symmetry at these two 
special values of the model parameter.

\section{Quantization of the system}

\subsection{State space of the model}

To describe the state space of the model, let us 
first remove from the theory the Lagrange multipliers 
$e$ and $v$ and their canonically conjugate momenta 
$p_e$ and $p_v$. To this end, we introduce gauge-fixing 
conditions $e - e_0 \approx 0$, $v - v_0 \approx 0$ 
for the primary constraints, where $e_0$ and $v_0$ 
are some constants. Using the notion of Dirac brackets, 
we can now define the quantum theory on the corresponding 
reduced phase space. 
Upon quantization, the odd variables $b^\pm_\mu$ become the 
fermionic creation-annihilation operators $\widehat{b}^\pm_\mu$ 
having the only nonzero anticommutators 
$[ \widehat{b}^-_\mu,\widehat{b}^+_\nu ]_+ = \eta_{\mu\nu}$.  
Then an arbitrary quantum state can be realized over the vacuum 
$|0\rangle$, defined as 
$\widehat{b}^-_\mu |0\rangle = 0$,
$\langle 0|0 \rangle = 1$:
\begin{equation}
\Psi(x) = \left( f(x) + {\cal F}^\mu(x) \widehat{b}_\mu^+ 
+ \frac{1}{2!} \varepsilon_{\mu\nu\lambda} {\tilde{\cal F}}^\mu(x) 
\widehat{b}^{+\nu} \widehat{b}^{+\lambda} 
+ \frac{1}{3!} \tilde{f}(x) \varepsilon_{\mu\nu\lambda}
\widehat{b}^{+\mu} \widehat{b}^{+\nu} \widehat{b}^{+\lambda} \right) 
|0\rangle.
\label{Psi}
\end{equation}
It is clear that Eq. (\ref{Psi}) means an expansion of the general
state vector into the complete set of eigenvectors of the fermion 
number operator $\widehat{\cal N}$. The coefficients of this 
expansion are some square-integrable functions of the space-time 
coordinates. The quantum parity and time-reversal transformations 
are generated by the antiunitary operators
\begin{equation}
U_P = V^0_+ V^1_- V^2_+, \qquad U_T = V^0_- V^1_+ V^2_+,
\end{equation}
\begin{equation}
U^\dagger_{P,T} = U_{P,T} \qquad U^2_{P,T} = - 1,
\end{equation}
where 
$V^\mu_\pm = \widehat{b}^{+\mu} \pm \widehat{b}^{-\mu}$,
as follows:
\begin{eqnarray}
&P,T : \Psi(x) \rightarrow \Psi^\prime(x^\prime_{P,T}) 
= U_{P,T} \Psi(x),& \\
&x^{\prime\mu}_P = (x^0,-x^1,x^2), \quad 
x^{\prime\mu}_T = (-x^0,x^1,x^2).
\end{eqnarray} 
In correspondence with classical relations (\ref{PT1}) 
we have
\begin{eqnarray}
U_P \widehat{b}^\pm_{0,2} U_P^{-1} &=& \widehat{b}^\mp_{0,2},
\qquad
U_P \widehat{b}^\pm_1 U_P^{-1} = - \widehat{b}^\mp_1,\\
U_T \widehat{b}^\pm_{1,2} U_T^{-1} &=& \widehat{b}^\mp_{1,2},
\qquad
U_T \widehat{b}^\pm_0 U_T^{-1} = - \widehat{b}^\mp_0.
\end{eqnarray}
We get that while acting on the general state $\Psi(x)$ 
these operators induce mutual transformation of scalar, 
$f(x) \leftrightarrow \tilde{f}(x)$, 
and vector, 
${\cal F}^\mu(x) \leftrightarrow \tilde{\cal F}^\mu(x)$,
fields.

\subsection{Physical subspace}

The physical states should be singled out by the quantum 
analogs of the remaining first class constraints: 
\begin{equation}
(P^2 + m^2) \Psi = 0
\qquad
\widehat{\chi} \Psi = 0,
\label{constr}
\end{equation}
where we assume that 
$P_\mu = -i \partial_\mu$. 
Note that the first class constraint corresponding to the
even nilpotent function $\chi$ admits no, even local, gauge 
condition, and so, the respective sector of the phase space 
can be quantized only by the Dirac method \cite{PR}. This 
peculiarity is caused by the homogeneous quadratic in 
Grassmann variables nature of $\chi$, due to which there 
exists no gauge constraint $\psi$ such that the bracket 
$\{\psi,\chi\}$ would be invertible. {}From Eq. (\ref{constr}) 
we see also that the scalar and vector functions from the state 
vector belong actually to the so-called Schwartz space, which 
is a rigged Hilbert space \cite{BLOT}.

Let us fix in the quantum operator $\widehat{\chi}$ the same 
ordering as in the corresponding classical constraint (\ref{sec}). 
This gives 
\begin{equation}
\widehat{\chi} = i\varepsilon_{\mu\nu\lambda} P^\mu 
\widehat{b}^{+\nu} \widehat{b}^{-\lambda} 
- q m ( \widehat{b}^+_\mu \widehat{b}^{-\mu} - 3/2).
\label{qu-chi}
\end{equation}
As a consequence of the quantum constraints 
(\ref{constr}), (\ref{qu-chi}), we find that 
\begin{equation}
f(x) = \tilde{f}(x) = 0,
\label{f}
\end{equation}
whereas the fields 
${\cal F}_\mu(x)$ and ${\tilde{\cal F}}_\mu(x)$
satisfy the equations  
\begin{equation}
i \varepsilon_{\mu\nu\lambda} P^\nu {\cal F}^\lambda 
- \frac{1}{2} q m {\cal F}_\mu = 0, \qquad
i \varepsilon_{\mu\nu\lambda} P^\nu {\tilde{\cal F}}^\lambda 
+ \frac{1}{2} q m {\tilde{\cal F}}_\mu = 0, \label{E}
\end{equation}
and
\begin{equation}
(P^2 + m^2) {\cal F}_\mu = (P^2 + m^2) {\tilde{\cal F}}_\mu = 0.
\label{KG}
\end{equation}
Due to the linear equations (\ref{E}) we have also 
\begin{equation}
P_\mu {\cal F}^\mu = P_\mu {\tilde{\cal F}}^\mu = 0
\label{trans}
\end{equation}
and
\begin{equation}
(P^2 + \frac{1}{4} q^2 m^2) {\cal F}_\mu 
= (P^2 + \frac{1}{4} q^2 m^2) {\tilde{\cal F}}_\mu = 0.
\label{q-KG}
\end{equation}
Comparing Eqs. (\ref{KG}) and (\ref{q-KG}), we see that the 
quantum constraints are consistent, and so, have nontrivial 
solutions if and only if $|q| = 2$. We have arrived at the 
same quantization condition which was obtained in the classical 
theory.

Note that we have obtained the transversality condition (\ref{trans})
for the vector fields while there was no corresponding constraint 
in the classical theory.
 
Putting $q = \epsilon 2$, $\epsilon = +$ or $-$, we finally see that 
the field ${\cal F}^\mu$ can be identified with the topologically
massive vector U(1) gauge field ${\cal F}_\epsilon^\mu$, whereas the
field ${\tilde{\cal F}}^\mu$ coincides with ${\cal F}^\mu_{-\epsilon}$.
This gives us the desirable $P$,$T$-invariant system \cite{top}.

Let us note here that the latter can be reformulated in terms
of the gauge fields through the duality relation 
$\varepsilon_{\mu\nu\lambda} {\cal F}_\epsilon^\lambda 
= F^\epsilon_{\mu\nu}
= \partial_\mu A^\epsilon_\nu - \partial_\nu A^\epsilon_\mu$.
In this case the corresponding basic equations are 
of the second order, and can thus be compared with 
equations of motion for another $P$ and $T$ conserving system 
-- gauge-non-invariant massive model. This one, the three-dimensional 
Proca theory, describes causally propagating massive field excitations 
of spin polarizations $+1$ and $-1$. 
So, the {\it kinematical} contents of the gauge-invariant and 
non-invariant cases are identical \cite{top,Bin}. 
However, our pseudoclassical model has led exactly to topologically 
massive gauge fields, but not to the Proca theory. The difference 
between these systems may appear {\it dynamically}, when the vector 
fields interact with matter fields. It is quite natural to expect
that interactions will affect these free systems in different ways.
In this respect, it would be worth investigating quantum symmetries 
of the three-dimensional gauge-non-invariant vector theory (see, for
example, Ref. \cite{Proca} where hidden parasupersymmetries of the 
four-dimensional Proca theory were analyzed) and conferring them with 
those we shall demonstrate in this paper. This could be done in the 
spirit of the present analysis with the help of a psedoclassical model 
corresponding to the Proca theory. Probably, such a model could be found 
as a result of modification of pseudoclassical models considered in Ref. 
\cite{pseudo}. 

If we choose another ordering prescription for the quantum counterpart
of the constraint function $\chi$, we would have the same operator 
but with the constant term $-3/2$ changed for $\alpha - 3/2$, where 
the constant $\alpha$ specifies the ordering \cite{CPV}. As a result, 
we would find that for $\alpha \neq 0, +3/2, -3/2$ under appropriate 
choice of the parameter $q$ (note in this case $|q| \neq 2$) we have 
as a solution of the quantum constraints only one field ${\cal F}_-^\mu$ 
or ${\cal F}_+^\mu$ satisfying the corresponding linear differential
equation. This would lead to the violation of the $P$ and $T$ symmetries 
at the quantum level. For $\alpha = +3/2$ (or $ q = 0$) or $\alpha = -3/2$ 
the physical states are respectively described by one scalar field $f(x)$ 
or $\tilde{f}(x)$, and for both these cases the discrete symmetries are 
broken. 

We see that the same values of the parameter $q$, $q = \pm 2$, which 
we have separated classically, turn out to be also special quantum
mechanically: for these the number of physical states is maximal,
so that the maximal global symmetry group can be realized on 
the physical state space, and only at $q = \pm 2$ parity and 
time-reversal symmetries are conserved. This result indicates 
that discrete and continuous global symmetries are profoundly 
connected.

\subsection{Scalar product and the field system}

To deal with the field system obtained upon quantization of the 
pseudoclassical model (\ref{Action})-(\ref{L}), let us consider 
average value of the constraint operator $\widehat{\chi}$ over 
an arbitrary state. First, let us investigate the structure of 
the scalar product on the state space. We find
\begin{equation}
\langle \Psi_2,\Psi_1 \rangle = \Psi^\dagger_2(x) \Psi_1(x) =
f^*_2(x) f_1(x) - \tilde{f}^*_2(x) \tilde{f}_1(x)
+ {\cal F}^*_{2\mu}(x) {\cal F}^\mu_1(x) 
- \tilde{\cal F}^*_{2\mu}(x) \tilde{\cal F}_1^\mu(x).
\end{equation}
{}From the last expression we see that the scalar product 
is indefinite in the doublets 
$\varphi = $ $f \choose \tilde{f}$ 
and 
$\Phi =$ ${\cal F} \choose \tilde{\cal F}$. 
Actually we have
\begin{equation}
\langle \Psi,\Psi \rangle 
= \bar{\varphi} \varphi + \bar{\Phi} \Phi,
\end{equation}
where 
$\bar{\varphi} = \varphi^\dagger \sigma_3$ 
and
$\bar{\Phi} = \Phi^\dagger \sigma_3$.
To have the norm of the state vectors defined from a
positive-definite scalar product, we should modify the 
metrics in the doublets $\varphi$ and $\Phi$ as follows:
\begin{equation}
\langle \Psi_2,\Psi_1 \rangle \rightarrow
\langle\langle \Psi_2,\Psi_1 \rangle\rangle
= \langle\Psi_2,\widehat\eta \Psi_1\rangle 
= \varphi_2^\dagger \varphi_1 + \Phi_2^\dagger \Phi_1,
\end{equation}
where the metric operator 
$\widehat\eta = (-1)^{\frac{1}{2}\widehat{\cal N}(\widehat{\cal N}-1)}$
is introduced. Remember that the discrete symmetry operators
$U_{P,T}$ are antiunitary with respect to the indefinite scalar
product $\langle . \rangle$. Using the relation
$U_{P,T}\widehat\eta = -\widehat\eta U_{P,T}$, it is easy to
verify that the operators $U_{P,T}$ are unitary with respect
to the modified scalar product $\langle\langle . \rangle\rangle$.

For the constraint operator $\widehat{\chi}$ we get the following
average value:
\begin{eqnarray}
\langle\langle \widehat{\chi} \rangle\rangle &\equiv& 
\langle\langle{\Psi}(x), \widehat{\chi} \Psi(x)\rangle\rangle 
\nonumber\\
&=& -i{\varepsilon^\alpha}_{\mu\beta} \left(
{\cal F}^*_\alpha P^\mu {\cal F}^\beta 
+ \tilde{\cal F}^*_\alpha P^\mu \tilde{\cal F}^\beta \right) 
+ \frac{1}{2} qm \left({\cal F}^*_\gamma {\cal F}^\gamma 
- \tilde{\cal F}^*_\gamma \tilde{\cal F}^\gamma \right) \nonumber\\
&& + \frac{3}{2}qm \left(f^*f - \tilde{f}^*\tilde{f}\right).
\end{eqnarray} 
We see that unphysical scalar fields $f(x)$ and $\tilde{f}(x)$
are completely decoupled from the physical sector, and so, we can
take into account the equations of motion (\ref{f}) for these fields
without changing physical contents of the theory, and put $q=\epsilon 2$. 
This gives
\begin{equation}
\langle\langle \widehat{\chi} \rangle\rangle_\epsilon
= \Phi^\dagger \left( PJ\otimes 1 + \epsilon m\cdot 1\otimes\sigma_3
\right) \Phi,
\label{4.19}
\end{equation}
where
$\Phi = ({\cal F}_\epsilon,{\cal F}_{-\epsilon})$
(in transposed form) and we use the usual conveniences 
with Pauli matrices and the generators 
$(J_\mu)^\alpha{}_\beta = -i\varepsilon^\alpha{}_{\mu\beta}$
in the vector representation of the 3d Lorentz group,
$[ J_\mu,J_\nu ] = - i\varepsilon_{\mu\nu\lambda}J^\lambda$,
$J_\mu J^\mu=-2$. The second factor in expressions with the
direct product, as in Eq. (\ref{4.19}), being either identity or
Pauli matrices, acts in the two-dimensional space labelled by
the index distinguishing spin $\epsilon$ and $-\epsilon$
components of the doublet $\Phi$,
whereas the first factor corresponds to its spin (vector) index.
The modified scalar product $\langle\langle . \rangle\rangle$ 
allows us to give the components of the doublets $\varphi$ and 
$\Phi$, and consequently, the spin states $+\epsilon$ and $-\epsilon$ 
equal treatment. However, note that there is still an indefiniteness 
due to the metric tensor $\eta_{\alpha\beta}$,
$\Phi^\dagger_2 \Phi_1 = 
{\cal F}^{*\alpha}_{2} \eta_{\alpha\beta} {\cal F}^{\beta}_{1}
+ \tilde{\cal F}^{*\alpha}_{2} \eta_{\alpha\beta} 
\tilde{\cal F}^{\beta}_{1}$.
The presence of $\eta_{\alpha\beta}$
guarantees the spinor part of the total angular momentum operator 
of the system to be a self-adjoint operator, 
$(\Phi_2^\dagger J_\mu \otimes 1 \Phi_1)^* 
= \Phi_1^\dagger J_\mu \otimes 1 \Phi_2$ 
\cite{CP}. But this indefiniteness only concerns pure gauge degrees 
of freedom present in the theory and does not actually play any 
role in our consideration.

Having incorporated the scalar fields into the theory, we provided
the completeness of the basis vectors of the total state space, 
expressed by the expansion (\ref{Psi}). The physical state space
is its subspace, obtained by eliminating the scalar fields. 
In this sense, $f(x)$ and $\tilde{f}(x)$ have actually been 
used as auxiliary fields. We get that on the physical 
subspace the space-time integral of the average value 
of the constraint operator coincides with the action 
${\cal A}$ (\ref{calA}): 
\begin{equation}
\int d^3x \langle\langle \widehat{\chi} \rangle\rangle_\epsilon 
= {\cal A} 
= \int d^3 x \Phi^\dagger(x) \left( PJ\otimes 1 
+ \epsilon m\cdot 1\otimes \sigma_3 \right) \Phi(x).
\label{system}
\end{equation}
Thus, the pseudoclassical model (\ref{Action})-(\ref{L}) leads 
to the $P,T$-invariant system of topologically massive vector 
U(1) gauge fields in a natural way. 

The corresponding procedure is reminiscent of that suggested 
in Ref. \cite{S} for constructing a string field theory action 
and subsequently developed in Ref. \cite{W}. There, a quantity
$A = \int d\mu \langle\Psi\vert\Omega\vert\Psi\rangle$, with
a BRST operator $\Omega$ singling out physical states and $d\mu$
being an integration measure, was regarded as a string field theory
action. In this, a scalar product $\langle\vert\vert\rangle$ was
proposed to provide hermiticity of the BRST operator. The 
underlying idea was originated from the observation that the
functional $A$ is extremal on the physical subspace: the
variational principle applied to the ``action'' $A$ results
in ``quantum equations of motion'' encoded in 
$\Omega\vert\Psi\rangle = 0$, and besides, it keeps
symmetries of the initial first-quantized theory.
In our case, we have an analogous construction, with 
the constraint operator $\widehat\chi$ instead of $\Omega$.  

\subsection{Quantum analogue of the Poisson-Dirac brackets}

In what follows, we put for brevity $\epsilon = +$, that corresponds 
to $q=2$. The case of $\epsilon = -$ ($q=-2$) can be achieved by 
obvious changes.

The quantum counterpart of the integrals of motion are
operators acting in the state space with the arbitrary 
state vector (\ref{Psi}). They form the following 
(super)algebra:
\begin{eqnarray}
&\left[ \widehat{\cal B}^+_+,\widehat{\cal B}^-_+ \right]_- = 
- 2 ( \widehat{\cal S} - \widehat{\cal R} ),
\label{B+B--}\\
&\left[ \widehat{\cal B}^\pm_+,\widehat{\cal S} \right]_- = 
\mp 2 \widehat{\cal B}^\pm_+, 
\qquad
\left[ \widehat{\cal B}^\pm_+,\widehat{\cal R} \right]_- =
\pm 2 \widehat{\cal B}^\pm_+,& 
\label{BpmRS}\\
&\left[ \widehat{\cal B}^+_+,\widehat{\cal B}^-_+ \right]_+ =
2\widehat{\cal C}_+, \qquad
\left[ \widehat{\cal B}^\pm_+,\widehat{\cal C}_+ \right]_- = 0,& 
\label{B+B-+}
\end{eqnarray}
where we have introduced the notations
\begin{equation}
\widehat{\cal R} = (1 + 2\widehat{\cal N}_0)
\widehat{\cal N}_\perp(2 - \widehat{\cal N}_\perp),
\qquad
\widehat{\cal C}_+ = (\widehat{\cal S} - \widehat{\cal R}) 
( 1 + 2\widehat{\cal N}_0 ). 
\label{R}
\end{equation}
Note that 
$(1 + 2\widehat{\cal N}_0)^2 = 1$.
Comparing Eqs. (\ref{b1})-(\ref{b3}) and Eqs. (\ref{B+B--})-(\ref{R}), 
we see that the (super)algebras of integrals of motion in the classical
and the quantum theories are essentially different. The reason of this
modification occurred at the quantum level is that the corresponding
operators are composite ones and, in particular, the integrals of
motion ${\cal B}^\pm_+$ are of the third order in (odd) spin variables
(we discuss this point in detail in the concluding Section).

\subsection{Quantum symmetry operators}

The generators of continuous global symmetries in the one-particle 
sector of the $P,T$-invariant system (\ref{system}) can be found
by averaging the quantum counterpart of the third order nilpotent 
integrals of motion taking place at $q = \epsilon 2$. 
For $\epsilon = +$ we have
\begin{equation}
{\cal Q}^\pm
= - \frac{1}{2}\langle\langle 
\widehat{\cal B}^\mp_+ \rangle\rangle
= \Phi^\dagger(x) Q^\pm \Phi(x),
\end{equation}
where the quantum mechanical nilpotent operators
\begin{equation}
Q^\pm = \frac{1}{4i} J^2_\pm \otimes \sigma_\pm
\end{equation}
realize mutual transformation of the physical states 
of spins $+1$ and $-1$. 
Here we use the notation
$\sigma_\pm = \sigma_1 \pm i\sigma_2$
and $J_\pm = J^{(1)} \pm iJ^{(2)}$,
$J^{(\alpha)} = J^\mu e^{(\alpha)}_\mu$.  
Commutation relation of these operators is 
\begin{equation}
[ Q^+,Q^- ]_- = \frac{1}{2} ( S - \Pi ),
\end{equation} 
where $S = J^{(0)}\otimes 1$ 
and $\Pi = J^{(0)} J^{(0)} \otimes \sigma_3$.
In this, $S$ is the spin operator corresponding to 
the average value of the quantum counterpart of the 
integral ${\cal S}$,
$\Phi^\dagger\; S\; \Phi 
= \langle\langle \widehat{\cal S} \rangle\rangle$,
and $\Pi$ is the operator associated with the projector 
onto the physical spin $\pm{1}$ states, that is the quantum 
counterpart of the integral of motion
${\cal R} = (1 + 2{\cal N}_0){\cal N}_\perp 
(2 - {\cal N}_\perp)$,
$\Phi^\dagger\; \Pi\; \Phi 
= \langle\langle \widehat{\cal R} \rangle\rangle$.
We have also 
\begin{equation}
\left[ Q^\pm,S \right]_- = \pm 2 Q^\pm, \qquad 
\left[ Q^\pm,\Pi \right]_- = \mp 2 Q^\pm.
\end{equation}
Besides, we find the anticommutator of the physical 
operators
\begin{equation}
[ Q^+,Q^- ]_+ = \frac{1}{2} (S\Pi - \Pi^2),  
\end{equation}
where 
\begin{equation}
S\Pi = J^{(0)}\otimes\sigma_3, \qquad
\Pi^2 = J^{(0)}J^{(0)}\otimes 1. 
\end{equation}
Taking into account that
$\Phi^\dagger\; S\Pi \; \Phi = \langle\langle 
\widehat{\cal S}(1 + \widehat{\cal N}_0) \rangle\rangle$
and
$\Phi^\dagger\; \Pi^2\; \Phi = \langle\langle
\widehat{\cal N}_\perp (2 - \widehat{\cal N}_\perp)\rangle\rangle$,
we finally see that the operators $Q^\pm$ reproduce exactly the 
(super)algebra of the quantum mechanical counterpart of the 
integrals ${\cal B}^\mp_+$ (\ref{B+B--})-(\ref{R}).

To obtain the above algebras of the quantum mechanical operators 
$Q^\pm$, we have used the properties of the triad and Pauli 
matrices, as well as the relations of the generators
$J^{(\alpha)}$, $\alpha = 0,1,2$, listed in the Appendix. 

As we have learned from Section 3, the nilpotent classical
constraint function $\chi$ of our model played the role of 
Hamiltonian for the spin variables. The operator 
\begin{equation}
D = PJ\otimes 1 + m\cdot 1\otimes \sigma_3
\end{equation} 
is its quantum analog obtained on the physical subspace 
after removing auxiliary scalar fields. It means that we can 
treat it as a quantum one-particle Hamiltonian, specifying 
simultaneously the physical state space, 
$D \Phi = 0$. 
The operators $Q^\pm$ generate symmetries of the operator $D$. 
Actually, we have
\begin{equation}
[ Q^\pm , D ]_- = \pm 2 Q^\pm (\sqrt{-P^2} - m) \approx 0,
\end{equation}
which means that $Q^\pm$ are quantum symmetry operators 
of the Hamiltonian $D$ \cite{GSHT}.

When considering the linear combinations 
\begin{equation}
Q_0 = \frac{1}{4} (S - \Pi), \quad 
Q_1 = \frac{1}{2} (Q^+ + Q^-), \quad  
Q_2 = \frac{i}{2} (Q^+ - Q^-),
\end{equation}
we obtain that the quantum physical operators $Q_\alpha$, 
$\alpha = 0,1,2$, form $su(1,1)$ algebra:
\begin{equation}
[ Q_\alpha,Q_\beta ]_- 
= -i \varepsilon_{\alpha\beta\gamma} Q^\gamma.
\end{equation} 
The generators $Q_\alpha$ and the Casimir operator
\begin{equation}
C = Q_\alpha \eta^{\alpha\beta} Q_\beta 
= \frac{3}{8} (S\Pi - \Pi^2) 
\end{equation}
of this algebra form also $s(2,1)$ superalgebra 
\cite{CrRit}, 
\begin{equation}
[ Q_\alpha,Q_\beta ]_+ 
= \eta_{\alpha\beta} \frac{2}{3} C, 
\qquad
[ Q_\alpha,C ]_- = 0,
\end{equation}
with even generator $C$ being different from the Hamiltonian
$D$.
   
The Casimir operator $C$ is related to the average value of
$\widehat{\cal C}_+$ from Eq. (\ref{R}),  
$\Phi^\dagger C \Phi 
= \frac{3}{8}\langle\langle\widehat{\cal C}_+\rangle\rangle$,
and it takes the value $C = -3/4$ on the physical subspace 
given by two square-integrable transversal vector fields
${\cal F}^\mu_+$, ${\cal F}^\mu_-$ 
carrying spins $-1$ and $+1$. 

We can also construct another combination of the physical operators
$S$ and $\Pi$, namely
\begin{equation}
U = \frac{1}{2}(S + \Pi),
\label{U}
\end{equation}
satisfying the relation
\begin{equation}
[ Q_\alpha,U ]_- = 0.
\end{equation}
The operator $U$ is the generator of global U(1) symmetry,
taking zero value on the physical states. 
Hence, the set of the physical operators 
$Q_\alpha$ and $U$
form ${\rm U(1,1)} = {\rm SU(1,1)}\times{\rm U(1)}$ group. 

We have thus revealed hidden U(1,1) symmetry and S(2,1) supersymmetry
of the $P$,$T$-invariant quantum system of topologically massive vector 
U(1) gauge fields at the one-particle level.

\section{Covariant form of the hidden (super)symmetries}

\subsection{Towards covariantization}

The hidden dynamical symmetry we have revealed in the previous Section 
leads to a non-standard super-extension of the (2+1)-dimensional 
Poincar\'e group. To show this we have to construct a covariant form 
of the above algebra relations. Actually, the quantities $S$ and $\Pi$, 
as well as their combination $Q_0$, 
\begin{equation}
Q_0 = \frac{1}{4} J^{(0)} \otimes 1 
- \frac{1}{4} J^{(0)} J^{(0)} \otimes \sigma_3, 
\label{Q0}
\end{equation}
are expressed as covariant (scalar) operators, 
whereas $Q_i$,
\begin{eqnarray}
Q_1 &=& \frac{1}{4i}\left[\left(J^{(1)}\right)^2 
- \left(J^{(2)}\right)^2\right]
\otimes \sigma_1 - \frac{1}{4i} \left[ J^{(1)}J^{(2)} + 
J^{(2)}J^{(1)} \right]
\otimes \sigma_2, \label{Q1} \\
&& \nonumber \\
Q_2 &=& - \frac{1}{4i}\left[J^{(1)}J^{(2)} + J^{(2)}J^{(1)}\right]
\otimes \sigma_1 - \frac{1}{4i} \left[\left(J^{(1)}\right)^2 - 
\left(J^{(2)}\right)^2\right] \otimes \sigma_2, \label{Q2} 
\end{eqnarray}
are given in terms of non-covariant quantities 
$J^{(i)}$, $i=1,2$. 

Taking into account that the hidden symmetry operators interchange
spins $+1$ and $-1$, it is natural to consider as a candidate for
their covariant form a rank-$2$ symmetric tensor operator
\begin{equation}
X_{\mu\nu} = X^{0}_{\mu \nu} + X^{\bot}_{\mu\nu}, \label{XXX}
\end{equation}
where
\begin{equation}
X^{0}_{\mu\nu} =  \e{0}_{\mu\nu} Q_0 , \label{X0}
\end{equation}
with $\e{0}_{\mu\nu}$ defined below, and $X^\perp_{\mu\nu}$ 
is a symmetric, transversal and traceless tensor:
\begin{equation}
X^\perp_{\mu\nu} = X^\perp_{\nu\mu}, \qquad
e^{(0)\mu} X^\perp_{\mu\nu} = 0, \qquad
\eta^{\mu\nu} X^\perp_{\mu\nu} = 0, 
\label{botX}
\end{equation}
being a quantum physical operator:
\begin{equation}
[ D , X_{\mu\nu} ]_- =
[ D , X^\bot_{\mu\nu} ]_- \approx 0, \label{DX}
\end{equation}
where the weak equality means the equality on the mass shell
$P^2 + m^2 \approx 0$. 

{}From the expressions of the quantum physical operators 
$Q_\alpha$ and $Q^\pm$ we see that it is convenient to 
introduce the quantities
\begin{equation}
e_\pm^{\mu \nu} = \left( e^{(1)\mu} \pm i e^{(2)\mu} \right) \cdot
\left( e^{(1)\nu} \pm i e^{(2)\nu} \right) \label{epm}
\end{equation}
and their linear combinations
\begin{eqnarray}
\e{1}^{\mu \nu} &=& \frac{1}{2\sqrt{2}} 
\left( e_+^{\mu \nu} + e_-^{\mu \nu} \right)
= \frac{1}{\sqrt{2}} 
\left( e^{(1)\mu} e^{(1)\nu} - e^{(2)\mu} e^{(2)\nu} \right),
\label{e1} \\
&& \nonumber \\
\e{2}^{\mu \nu} &=& \frac{i}{2\sqrt{2}} 
\left( e_+^{\mu \nu} - e_-^{\mu \nu} \right)
= - \frac{1}{\sqrt{2}} \left( e^{(1)\mu} e^{(2)\nu} 
+ e^{(2)\mu} e^{(1)\nu} \right), \label{e2}
\end{eqnarray}
together with the obvious covariant construction of the form
\begin{equation}
\e{0}^{\mu \nu} = i e^{(0)\mu} e^{(0)\nu} \label{e0}
\end{equation}
providing projection of the tensor $X_{\mu\nu}$ onto 
the set of non-covariant quantum symmetry operators:
\begin{equation}
\e{\alpha}^{\mu \nu} X_{\mu \nu} = Q^\alpha = 
\eta^{\alpha\beta} Q_\beta, \qquad \alpha, \beta = 0,1,2.
\label{eX}
\end{equation}

We find the solution to the equalities 
(\ref{botX})-(\ref{DX}),(\ref{eX}) in the form
\begin{equation}
X^{\bot}_{\mu\nu} = \frac{1}{4i} \A{1}_{\mu \nu} \otimes \sigma_1
+ \frac{1}{4i} \A{2}_{\mu \nu} \otimes \sigma_2, \label{Xbb}
\end{equation}
where the rank-$2$ tensors $\A{1}_{\mu \nu}$ and $\A{2}_{\mu \nu}$
are given by the expressions
\begin{equation}
\A{1}_{\mu \nu} = \frac{1}{\sqrt{2}}  
\left(\, r_\mu r_\nu - s_\mu s_\nu \, \right), 
\qquad
\A{2}_{\mu \nu} = \frac{1}{\sqrt{2}}
\left(\, r_\mu s_\nu + s_\mu r_\nu \, \right), 
\label{A1A2} 
\end{equation}
with
\begin{equation}
r_\mu = \pi_{\mu\nu} J^\nu \equiv J^\bot_\mu 
\equiv J_\mu + e^{(0)}_\mu J^{(0)}, \qquad
s_\mu = \varepsilon_{\mu\alpha\beta} e^{(0)\alpha} J^\beta . 
\label{rs}
\end{equation}
Here $\pi_{\mu\nu}$ is the quantum counterpart of the tensor
introduced by Eq. (\ref{pi}).

Using the properties of $r_\mu$ and $s_\mu$, described in the
Appendix, we obtain that the square of the tensor $X_{\mu\nu}$ 
is the Casimir operator $C$:
\begin{equation}
X_{\mu\nu} \cdot X^{\mu\nu} = 
X^0_{\mu\nu} \cdot X_0^{\mu\nu} +
X^\bot_{\mu\nu} \cdot X_\bot^{\mu\nu} = C. 
\label{X2=C}
\end{equation}

Let us introduce the notations
\begin{equation}
{\cal G}^{\mu\nu \vert \rho\sigma} =
\e{\alpha}^{\mu\nu} \eta_{\alpha\beta} \e{\beta}^{\rho\sigma}, 
\qquad
{\cal E}^{\mu\nu \vert \rho\sigma \vert \lambda\tau} =
\varepsilon_{\alpha\beta\gamma} 
\e{\alpha}^{\mu\nu} \e{\beta}^{\rho\sigma} 
\e{\gamma}^{\lambda\tau}.  
\label{calG1E1}
\end{equation}
The properties of the quantities $\e{\alpha}_{\mu\nu}$ 
and of the tensors 
${\cal G}_{\mu\nu\vert\rho\sigma}$ 
and 
${\cal E}_{\mu\nu\vert\rho\sigma\vert\lambda\tau}$ 
are listed in the Appendix.
Using these properties and the relation between the Casimir 
and $Q_0$ operators, we get that the tensor operator 
$X^\bot_{\mu\nu}$ satisfies the equation
\begin{equation}
X^\bot_{\mu\nu} X^\bot_{\rho\sigma} = \frac{1}{2} \left(\,
\pi_{\mu\nu} \pi_{\sigma\rho} - \pi_{\mu\sigma} \pi_{\nu\rho} - 
\pi_{\mu\rho} \pi_{\nu\sigma} \, \right) Q_0^2 
+ \frac{i}{4} \left(\, 
\pi_{\mu\sigma} \varepsilon_{\nu\rho\lambda} + 
\pi_{\nu\rho} \varepsilon_{\mu\sigma\lambda} \, 
\right) e^{(0)\lambda} Q_0. \label{XbXb3}
\end{equation}
The last equality leads finally to the symmetry algebra 
relation for the tensor operator $X_{\mu\nu}$: 
\begin{equation}
X_{\mu\nu} X_{\rho\sigma} = {\cal G}_{\mu\nu \vert \rho\sigma} 
\cdot \frac{1}{3} C 
\,-\, \frac{i}{2} {\cal E}_{\mu\nu \vert \rho\sigma \vert \lambda\tau} 
X^{\lambda\tau} . 
\label{XXF}
\end{equation}
In addition to Eq. (\ref{XXF}) we have the commutation relations
\begin{equation}
[X_{\mu\nu},C]_- = 0, \qquad
[X_{\mu\nu},U]_- = 0, \label{XCU}
\end{equation}
where the operator $U$ is expressed by Eq. (\ref{U}).

It is worthwhile seeing that, as required by tensor nature of 
the hidden symmetry generators, the covariantization is achieved 
by means of a kind of the bi-vector mapping, provided that the
quantities $\e{\alpha}_a$, $a = (\mu\mu^\prime)$, play the same 
role in the corresponding bi-vector space as the components of 
the complete oriented triad $e^{(\alpha)}_\mu$ do in the 
three-dimensional Minkowski space-time. This actually
originates our notations of Eq. (\ref{calG1E1}) and consequent
properties, so that ${\cal G}_{ab}$ and ${\cal E}_{abc}$
turn out to be the metric and totally antisymmetric tensors
of the bi-vector space.  

Note that one can introduce a complex vector $a_\mu$ related to the
operators $r_\mu$ and $s_\mu$ as $2 a_\mu = r_\mu + i s_\mu$,
$a_\mu a^\mu = 0$. Then the tensor operator $X^\bot_{\mu\nu}$
is represented in the form
$X^\bot_{\mu\nu} = \frac{1}{\sqrt{2}i} 
\theta^*_\mu \sigma_1 \theta_\nu$, 
where the $2 \times 2$ block-matrix $\theta_\mu$ 
is given by the expression
$\theta_\mu = {\rm diag}\;(a_\mu , a^*_\mu)$. 
Obviously, the same (super)algebras of the dynamical symmetry operators 
can be obtained in terms of this representation as well. Probably,
for some particular problems the use of the complex-valued vector
operator $a_\mu$ would be more appropriate than of its real and
imaginary parts $r_\mu$ and $s_\mu$, while the latter seem to be
quite sufficient for the present analysis. 

\subsection{Non-standard super-extension of the Poincar\'e group}

We have now the following set of covariant operators
being the generators of the full dynamical symmetry algebra: 
$X_{\mu\nu}$ -- generators of the hidden symmetries just revealed, 
$P_\mu$ -- energy-momentum operator, $M_\mu$ -- total angular momentum
operator explicitly given by the expression
\begin{equation}
M_\mu =  - \varepsilon_{\mu\nu\lambda} x^\nu P^\lambda \cdot 
1 \otimes 1 \,+\, J_\mu \otimes 1. \label{l1}
\end{equation}
One can easily get that the above operators are integrals of motion.
Actually, the commutator of $X_{\mu\nu}$ with the Hamiltonian 
$D$ disappears in a weak sense on the surface defined by the 
constraint $P^2 + m^2 \approx 0$, while the generators $P_\mu$ 
and $M_\mu$ strongly commute with $D$: 
\begin{equation}
[ D , X_{\mu\nu} ]_- \approx 0, \qquad
[ D , M_\mu ]_- = 0, \qquad
[ D , P_\mu ]_- = 0.
\label{Dcom}
\end{equation}  

Nonzero (anti)commutation relations of these operators are of the form:
\begin{eqnarray}
&\left[ M_\mu , P_\nu \right]_-& = 
-i \varepsilon_{\mu\nu\lambda} P^\lambda,
\label{MP}\\
&\left[ M_\mu , M_\nu \right]_-& = 
-i \varepsilon_{\mu\nu\lambda} M^\lambda,
\label{ll}\\
&\left[ X_{\mu\nu} , X_{\rho\sigma} \right]_+& = 
{\cal G}_{\mu\nu \vert \rho\sigma} \cdot \frac{2}{3} C,
\label{XX+}\\
&\left[ X_{\mu\nu} , X_{\rho\sigma} \right]_-& = 
-i {\cal E}_{\mu\nu \vert \rho\sigma \vert \lambda\tau} X^{\lambda\tau},
\label{XX-}\\
&\left[ M_\mu , X_{\rho\sigma} \right]_-& =
-i \varepsilon_{\mu\rho\lambda} {X^\lambda}_\sigma 
-i \varepsilon_{\mu\sigma\lambda} {X_\rho}^\lambda. 
\label{lX}
\end{eqnarray} 
Hence, the physical operators $P_\mu$, $M_\mu$ and $X_{\mu\nu}$ 
complete the set of generators of the superextended Poincar\'e group 
$\rm ISO(2,1|2,1)$. The Casimir operators of this supergroup 
are $P^2$ and the {\it superspin} 
\begin{equation}
\Sigma = e^{(0)\mu} M_\mu + 2 \e{0}^{\mu\nu} X_{\mu\nu}.
\end{equation}
{}From the explicit forms of the total angular momentum operator
$M_\mu$ and the hidden (super)symmetry generators $X_{\mu\nu}$ we
find
\begin{equation}
\Sigma = \frac{1}{2}(S + \Pi) = U.
\end{equation}
We get that the physical operator $U$ has the sense of 
the superspin of the system. The eigenvalues of the superspin
$\Sigma$ are given by the set of numbers $(-1,0,0,0,0,1)$. 
As we noted in the preceding Section, the operator $U$ takes zero
value in the physical subspace. Therefore, we gain that the physical 
states are the eigenstates of the superspin operator with zero eigenvalue. 
The same result can be seen by expressing the operator $C$ through the 
superspin as a quadratic function of the superspin:
$C = \frac{3}{4} ( \Sigma^2 - J^{(0)}J^{(0)}\otimes 1 )$. 
Consequently, the one-particle states of the quantum $P$,$T$-invariant 
system of topologically massive vector U(1) gauge fields realize an 
irreducible representation of the supergroup $\rm ISO(2,1|2,1)$ labelled 
by the zero eigenvalue of the superspin. Similar properties have been 
elucidated for the double fermion system \cite{GPS}, which is also 
considered to be relevant to high-temperature superconductivity 
\cite{T-inv-mod}.

\section{Concluding remarks}

In this paper, with the help of the proposed pseudoclassical model
(\ref{Action})-(\ref{L}) we have uncovered a rich set of hidden
symmetries of the $P$,$T$-invariant system of topologically massive 
vector U(1) gauge fields. 

Let us stress once more on the difference between the (super)algebras
formed by the integrals of motion at the classical and the quantum
levels. In the classical theory we have the set of quadratic
(in independent odd variables) integrals of motion and two
additional third order quantities conserved at the special
values of the model parameter. These two integrals of motion
together with the ``Hamiltonian'' ${\cal H}_+$ formed $N=2$ 
supersymmetry algebra (\ref{b2}) with respect to the Poisson-Dirac 
brackets, and the system reproduced this superalgebra at the quantum
level: we have the relations (\ref{B+B-+}) where the operator
$\widehat{\cal C}_+$ plays the role of the quantum counterpart 
of the ``Hamiltonian'' ${\cal H}_+$ (the corresponding modification 
is actually due to the compositeness of the integrals of motion).  
At the quantum level we have also $su(1,1)$ algebra, given
by Eqs. (\ref{B+B--})-(\ref{BpmRS}) with respect to commutators. 
This symmetry algebra can be reproduced at the classical level 
only partially: to the commutator of the quantum counterparts
of the third order integrals of motion corresponds the relation
(\ref{b1}) defined with respect to ordinary multiplication, so
that we lose the usual correspondence between commutators and
canonical brackets. To see the reason for such a breakdown of
the standard quantization prescription \cite{Dirac},
$\{\,,\,\} \rightarrow [\,,\,]/i\hbar$,
with respect to the hidden symmetry algebras, one has to 
reconstruct the spin variables in physical units, 
$\widehat b^\pm_\mu \sim \hbar^{1/2}$, so that
$\widehat{\cal B}^\pm_+ \sim \hbar^{3/2}$. 
Now it is clear that (anti)commutators of the operators 
$\widehat{\cal B}^\pm_+$, divided by $i\hbar$, vanish
in the classical limit, 
$[\,,\,]/i\hbar \rightarrow 0$ as $\hbar \rightarrow 0$. 
Actually, this observation just corresponds to the well-known 
fact that the terms of order $\hbar^{2+\kappa}$, $\kappa \ge 0$, 
have no classical analog in ordinary (without Grassmann variables) 
classical mechanics.

This situation is different from that one realized for the case 
of planar fermions \cite{GPS}. In the latter, all the integrals 
of motion forming hidden dynamical symmetry group are quadratic 
in odd variables. Besides, there is an odd first order integral 
of motion in the double fermion system, which gives a possibility 
to change Grassmann parities of the integrals of motion simply
multiplying them by this quantity. All this allows one to have
one and the same (super)symmetry algebras in the corresponding
classical and quantum theories. As we have seen above, the
situation considered in this paper is essentially different,
and so, quantum symmetries are reproduced at the classical
level only in part. Nevertheless, exactly the quantum counterparts  
of the third order integrals of motion give us finally the set
of revealed quantum symmetries.

It is interesting to compare the supersymmetry we have revealed
in this paper with the BRST and anti-BRST type supersymmetry,
obtained in Ref. \cite{L-gauge} for the non-Abelian Chern-Simons
theory and shown in Ref. \cite{DR} to be forming $\rm IOSp(3\vert 2)$ 
supergroup. The latter has been proven only for Landau gauge, provided 
that namely in this particular gauge ghost and vector field sectors are 
respectively coupled. The hidden supersymmetry we have elucidated for 
$P,T$-invariant Abelian Chern-Simons theory is not related to any 
gauge-choice, it is a {\it true supersymmetry} leading to a 
nontrivial super-extension of the Poincar\'e group.  

The pseudoclassical model we have proposed for $P,T$-invariant
system of topologically massive vector U(1) gauge fields turned 
out itself to be very interesting. It has revealed the quantization 
of the parameter $q$ and nontrivial `superposition' of the discrete 
($P$ and $T$) and continuous (hidden U(1,1) and S(2,1)) 
(super)symmetries. These hidden continuous global symmetries
are quite nontrivial since the corresponding generators act
not only on spin $+1$ and $-1$ states as the whole, but they 
transform also components of the fields and, moreover, due 
to the dependence on $P_\mu$, the symmetry generators act 
nontrivially on the space-time coordinates. 

A principal problem to be investigated developing these results 
is to obtain the quantum field analog of the hidden (super)symmetry 
generators. It is quite natural to expect that they should be 
generators of the corresponding field symmetry transformations. 
Having such an interpretation, one might further analyze
systems with $P,T$-invariant matter coupling and study 
what happens with the revealed hidden (super)symmetries.

\vskip1cm
{\bf Acknowledgements}
\vskip5mm
The work of Kh.N. has been supported by the Alexander von Humboldt
Fellowship and by the European Commission TMR programme 
ERBFMRX--CT96--0045 and ERBFMRX--CT96--0090.
M.P. thanks Prof. W. R\"uhl and the University of Kaiserslautern, 
where a part of this work has been realized, for kind hospitality.

\appendix

\section{Appendix}

\subsection{Generators of the 3d Lorentz group}

The projections of the generators of the 3d Lorentz group
onto the triad, as introduced in Section 4.5, satisfy the 
relations
\begin{eqnarray}
\left[ J^{(\alpha)},J^{(\beta)} \right]_- &=& 
-i\varepsilon^{\alpha\beta\gamma}J_{(\gamma)}, 
\label{B1}\\
\left[ J_\pm^2,J^{(0)} \right]_- &=& \pm 2 J_\pm^2, 
\label{B2}\\
\left[ J_+^2,J_-^2 \right]_- &=& - 4J^{(0)},
\label{B3}\\
\left[ J_+^2,J_-^2 \right]_+ &=& 4 J^{(0)} J^{(0)},   
\label{B4}
\end{eqnarray}
and 
\begin{equation}
\left(J^{(0)}\right)^{2k} = J^{(0)} J^{(0)}, \qquad
\left(J^{(0)}\right)^{2k+1} = J^{(0)}
\label{B5}
\end{equation}
for any positive integer $k$.

\subsection{Vector operators $r_\mu$ and $s_\mu$ and their combinations}

The vectors $r_\mu$ and $s_\mu$ have the properties
\begin{equation}
e^{(i)\mu} r_\mu = J^{(i)}, \quad 
e^{(i)\mu} s_\mu = \delta^{i2} J^{(1)} - \delta^{i1} J^{(2)},
\quad e^{(0)\mu} r_\mu = e^{(0)\mu} s_\mu = 0, \label{rsJ}
\end{equation}
from which we get the transversality condition
\begin{equation}
e^{(0)\mu} \A{i}_{\mu\nu} = 0, \qquad i =1,2. \label{e0Ai}
\end{equation}
Besides, we obtain the equalities
\begin{eqnarray}
r_\mu r^\mu &=& s_\mu s^\mu = - 2 + J^{(0)} J^{(0)}, \label{rrss} \\
r_\mu s^\mu &=& - s_\mu r^\mu = - i J^{(0)} , \label{rssr} 
\end{eqnarray}
which provide tracelessness of the rank-2 tensors $\A{i}_{\mu\nu}$:
\begin{equation}
\A{i}^\mu_{\mu} = 0, \qquad i =1,2. \label{Amm}
\end{equation}

The following relations are useful to clarify properties
of the tensors $\A{i}_{\mu\nu}$: 
\begin{eqnarray}
&[ r_\mu, r_\nu ]_- = 
[ s_\mu, s_\nu ]_- = 
i \varepsilon_{\mu\nu\lambda} e^{(0)\lambda} J^{(0)}, \qquad
[ r_\mu, s_\nu ]_- = 
- i \pi_{\mu\nu} J^{(0)},& \label{com1} \\
&&\nonumber\\
&[ J^{(0)},r_\mu ]_- = i s_\mu, \qquad
[ J^{(0)},s_\mu ]_- = - i r_\mu,& \label{com2}\\
&&\nonumber\\
&\varepsilon_{\mu\alpha\beta} e^{(0)\alpha} r^\beta = s_\mu, \qquad
\varepsilon_{\mu\alpha\beta} e^{(0)\alpha} s^\beta = - r_\mu.& 
\label{p4}
\end{eqnarray}
It follows from Eqs. (\ref{com2}) that
\begin{equation}
[ J^{(0)}, \A{1}_{\mu\nu} ]_- = 
2 i \A{2}_{\mu\nu}, \qquad 
[ J^{(0)}, \A{2}_{\mu\nu} ]_- = 
- 2 i \A{1}_{\mu\nu}. \label{com4}
\end{equation}
The last equalities are necessary to prove that $X_{\mu\nu}$
is a physical operator.

Using the properties of the operators $r_\mu$ and $s_\mu$, 
we obtain the relations
\begin{eqnarray}
&\A{1}_{\mu\nu} \cdot \A{1}^{\mu\nu} = 
\A{2}_{\mu\nu} \cdot \A{2}^{\mu\nu} =  2 J^{(0)} J^{(0)},& 
\label{AiAi} \\
&& \nonumber \\
&\A{1}_{\mu\nu} \cdot \A{2}^{\mu\nu} = 
- \A{2}_{\mu\nu} \cdot \A{1}^{\mu\nu} = 2 i J^{(0)},& 
\label{AiAj} 
\end{eqnarray}
which help us to relate the operator $X_{\mu\nu}$ with the
Casimir operator.

The operators $\A{i}_{\mu\nu}$ fulfil also the equalities
\begin{eqnarray}
J^{(0)} J^{(0)} \A{i}_{\mu\nu} &=& \A{i}_{\mu\nu} 
= \A{i}_{\mu\nu} J^{(0)} J^{(0)}, \qquad i = 1,2, 
\label{JJAi} \\
&& \nonumber \\
i J^{(0)} \A{1}_{\mu\nu} &=& - \A{2}_{\mu\nu} 
= - i \A{1}_{\mu\nu} J^{(0)},  \label{JA1} \\
&& \nonumber \\
i J^{(0)} \A{2}_{\mu\nu} &=& \A{1}_{\mu\nu} 
= - i \A{2}_{\mu\nu} J^{(0)},  \label{JA2} 
\end{eqnarray}
which are necessary to obtain the (super)algebra of the operators
$X_{\mu\nu}$. 

\subsection{Structure functions of symmetry algebras}

The list of useful properties of the quantities 
$\e{\alpha}^{\mu\nu}$ is
\begin{eqnarray}
\e{\alpha}^{\mu\nu} \cdot \e{\beta}_{\mu\nu} 
&=& \eta^{\alpha\beta},
\label{prop}\\
&&\nonumber\\
\varepsilon_{0ij} \e{i}_{\mu\nu} \e{j}_{\rho\sigma} 
&=& \frac{1}{2}
\left(\, \pi_{\mu\sigma} \varepsilon_{\nu\rho\tau} +
\pi_{\nu\rho} \varepsilon_{\mu\sigma\tau} \, \right) e^{(0)\tau},
\label{p1} \\
&& \nonumber \\
\varepsilon_{\alpha\beta\gamma} \e{\alpha}_{\mu\nu}
\e{\beta}_{\rho\sigma} \e{\gamma}_{\lambda\tau} \equiv 
{\cal E}_{\mu\nu \vert \rho\sigma \vert \lambda\tau} &=& 
\frac{1}{2}
\e{0}_{\lambda\tau} \left(\, \pi_{\mu\sigma} \varepsilon_{\nu\rho\alpha}
+ \pi_{\nu\rho} \varepsilon_{\mu\sigma\alpha} \, \right) e^{(0)\alpha}
\nonumber \\ 
&&\nonumber \\
&& + \frac{1}{2}
\e{0}_{\mu\nu} \left(\, \pi_{\rho\tau} \varepsilon_{\sigma\lambda\alpha}
+ \pi_{\sigma\lambda} \varepsilon_{\rho\tau\alpha} \, \right) e^{(0)\alpha}
\nonumber \\
&&\nonumber\\
&& \;\; + \frac{1}{2}
\e{0}_{\rho\sigma} \left(\, \pi_{\mu\tau} \varepsilon_{\lambda\nu\alpha}
+ \pi_{\nu\lambda} \varepsilon_{\tau\mu\alpha} \, \right) e^{(0)\alpha}\, ,
\label{Eps1} \\
&& \nonumber \\
\e{\alpha}^{\mu\nu} \eta_{\alpha\beta} \e{\beta}^{\rho\sigma} \equiv
{\cal G}^{\mu\nu \vert \rho\sigma} &=& - \e{0}^{\mu\nu} \e{0}^{\rho\sigma}
+ \frac{1}{2} \left(\, \pi^{\mu\rho} \pi^{\nu\sigma} + 
\pi^{\mu\sigma} \pi^{\nu\rho} - \pi^{\mu\nu} \pi^{\rho\sigma} \,\right).
\label{calG2}
\end{eqnarray}
{}From the above relations we see that
\begin{equation}
{\cal G}^{\mu\nu \vert \rho\sigma}  
X_{\rho\sigma} =  \e{0}^{\mu\nu} Q_0 +  X_{\bot}^{\mu\nu} = X^{\mu\nu},
\label{eeX}
\end{equation}
\begin{eqnarray}
{\cal E}^{\mu\nu\vert\rho\sigma\vert\lambda\tau} X_{\lambda\tau} 
&=& - \frac{1}{2}
\left(\, \pi^{\mu\sigma} \varepsilon^{\nu\rho\tau} + \pi^{\nu\rho}
\varepsilon^{\mu\sigma\tau} \, \right) e^{(0)}_\tau Q_0 \nonumber \\
&& \nonumber \\
&& - e^{(0)\mu} e^{(0)\nu} Q_0 X_\bot^{\rho\sigma} 
 - e^{(0)\rho} e^{(0)\sigma} X_\bot^{\mu\nu} Q_0\, . \label{eeeX} 
\end{eqnarray}
Then, taking into account the symmetry algebra relation
\begin{equation}
Q^\alpha Q^\beta = \eta^{\alpha\beta} \frac{1}{3} \, C - 
\frac{i}{2} \, {\varepsilon^{\alpha\beta}}_\gamma Q^\gamma, 
\label{QaQb}
\end{equation}
we obtain a covariant equation
\begin{equation}
X_\bot^{\mu\nu} X_\bot^{\rho\sigma} = \frac{1}{6} \left(\,
\pi^{\mu\sigma} \pi^{\nu\rho} 
- \varepsilon^{\mu\sigma\lambda} e^{(0)}_\lambda
\varepsilon^{\nu\rho\tau} e^{(0)}_\tau \, \right)\cdot C 
+ \frac{i}{4} \left(\, 
\pi^{\mu\sigma} \varepsilon^{\nu\rho\lambda} + 
\pi^{\nu\rho} \varepsilon^{\mu\sigma\lambda} \, 
\right) e^{(0)}_\lambda \cdot Q_0. \label{XbXb}
\end{equation}
Finally, the equality
\begin{equation}
\varepsilon_{\mu\sigma\lambda} e^{(0)\lambda}
\varepsilon_{\rho\nu\tau} e^{(0)\tau} 
= \pi_{\mu\rho} \pi_{\sigma\nu} - \pi_{\mu\nu} \pi_{\sigma\rho}
\label{eePP}
\end{equation}
is implicated to write the structure functions of the 
hidden (super)symmetry algebra in a more appropriate
form, given by Eqs. (\ref{prop})-(\ref{calG2}).

The tensor ${\cal E}$ has useful properties
\begin{equation}
\e{\alpha}^{\mu\nu} \, \e{\beta}^{\rho\sigma} \, 
{{\cal E}_{\mu\nu \vert \rho\sigma \vert}}^{\lambda\tau} = 
\varepsilon^{\alpha\beta\gamma} e_{[\gamma]}^{\lambda\tau}\, , \qquad
e_{[\gamma]}^{\lambda\tau} \equiv \eta_{\gamma\delta} \, 
\e{\delta}^{\lambda\tau} \, . \label{Eps2}
\end{equation}
Besides, the tensor ${\cal G}_{\mu\nu \vert \rho\sigma}$ is symmetric
over the pair indices: ${\cal G}_{ab} = {\cal G}_{ba}$, where $a = (\mu\nu)$
$b= (\rho\sigma)$, while the tensor ${\cal E}_{abc}$ with $a = (\mu\nu)$,
$b = (\rho\sigma)$, $c = (\lambda\tau)$ is completely antisymmetric over
pair indices. It is remarkable that basic properties of the tensors 
${\cal E}$ and ${\cal G}$ are the same as of the totally
antisymmetric tensor $\varepsilon_{\mu\nu\lambda}$ and of 
the metric tensor $\eta_{\mu\nu}$ of the 3d Minkowski 
space-time. Actually, we have
\begin{eqnarray}
{\cal E}_{abc} {\cal E}^{abc} &=& - d! = -6, \label{1.} \\
{\cal E}_{acd} {{\cal E}_b}^{cd} &=& - 2 {\cal G}_{ab}, \label{2.} \\
{\cal E}_{abf} {{\cal E}^f}_{cd} &=& {\cal G}_{ad} {\cal G}_{bc}
- {\cal G}_{ac} {\cal G}_{bd}, \label{3.} 
\end{eqnarray}
and 
\begin{equation}
{\cal G}_{ab} = {\cal G}_{ba}, \qquad 
{\cal G}_{ac} {\cal G}^{cb} = {{\cal G}_a}^b, \qquad 
{\cal G}_{ab} {\cal G}^{ab} = d = 3, \label{4.}
\end{equation}
where $d = 3$ is the space-time dimension.

\end{document}